\documentclass[eprint, twocolumn]{revtex4-2}
\usepackage{bbm}
\usepackage{natbib}
\usepackage{graphics}
\usepackage{amsmath}
\usepackage{mathrsfs}
\usepackage{theorem}
\usepackage{float}
\usepackage{hyperref}
\usepackage{rotating}
\usepackage{amssymb}
\usepackage[english]{babel}
\usepackage{color}
\usepackage{fancybox}
\newcommand{\ket}[1]{|#1\rangle}

\newcommand{\be}{\begin{eqnarray}}
\newcommand{\ee}{\end{eqnarray}}

\begin{document}

\title{Exact approach to double- and triple-excitation bound state in open quantum systems}
\author{X. S. Sun $^{1}$}
\author{H. T. Cui $^{1}$}
\email{cuiht01335@aliyun.com}
\author{Y. A. Yan $^{1}$}
\email{yunan@ldu.edu.cn}
\author{H. Z. Shen $^{2}$}
\email{shenhz458@nenu.edu.cn}
\affiliation{$^1$ School of Physics and Optoelectronic Engineering \& Institute of Theoretical Physics, Ludong University, Yantai 264025, China}
\affiliation{$^2$ Center for  Quantum Sciences, Northeast Normal University, Changchun 130024, China}
\date{\today}

\begin{abstract}
The  determination of multiple-excitation bound state in open quantum systems is of broad interest in quantum physics. While the emergence of double- or triple-excitation bound state  has been demonstrated in certain specific situations, their universality and exact solvability remain elusive. In this work, we present an exact  approach to the double- or triple-excitation  dynamics  for interacting quantum emitters coupled to environment by deriving a closed equation.  By this exact equation, we identify the critical condition for the emergence of double- or triple-excitation bound state, as well as its relationship to the single-excitation one.   Our study not only  clearly demonstrates the universality of double- or triple-excitation bound state in open quantum emitter systems, but also reveals the possibility of mediating inter-emitter interactions via multiple-excitation bound states.
\end{abstract}

\maketitle

\section{introduction}

The interplay of non-Markovianity,  system-environment bound state and inter-excitations interaction can give rise to the  intricate dynamics in open quantum systems \cite{xu26, va17}. Disentangling that interplay calls for exact solutions, yet an exact treatment  imposes a great challenge in theory since the system and its environment become strongly entangled. To address this difficulty,  a variety of methods  have been developed to capture the key properties of open dynamics, such as  the master equation  approach \cite{hp02} and other perturbative schemes \cite{xu26,va17}. Nevertheless, it  is of vital importance to develop the exactly solvable situation in order to test the approximation against exact result and assess the validity regime of these approaches.

The single excitation subspace is a special exactly solvable situation, in which only one excitation is present and conserved in the whole system. Resultantly, a closed equation can be founded, which provides the complete perspective on excitation dynamics in system.  A particular result of this equation is determination of single-excitation bound state (SBS), which characterizes  robustness of single excitation against the dissipation induced by its coupling to the environment  \cite{john91,kofman94}.  The SBS  has been utilized to manage the correlation in systems \cite{calojo}, and to control the status of quantum systems by engineering the environment\cite{douglas15,liu17}. 

However, the single-excitation case excludes the excitation interaction, which is vital for the emergence of exotic dynamics in open quantum systems. In particular, it is expected that the multiple-excitation bound states may emerge when the excitation interaction occur.  Recently, numerous efforts have been devoted to addressing  this issue, both experimentally or theoretically. In quantum optics, the multi-photon bound states have been observed experimentally \cite{firstenberg,liang18, cantu20}. Through manipulating the photon-atom interaction, either effective repulsive or attractive interactions between photons can be achieved.  The emergence of photon bound state is responsible for a variety of exotic phenomena, such as stable Rabi oscillations\cite{talukdar}, long-range four-body interaction \cite{wang24} and two-body mobility edges \cite{wang25}. In optical lattices, bound atom pairs has been realised  experimentally \cite{winkler06,folling07}, by exploiting the lattice band structure and repulsion between constituents\cite{winkler06}.  Moreover, the bound atom pairs can illustrate exotic features, such as Anderson localization  \cite{dufour12} and the correlated Bloch oscillations \cite{preiss15, cai21}.

In contrast to these practical studies, a number of theoretical works have addressed  the formation of two-body or three-body bound states in general settings. For instance,  Ref. \cite{calojo} studies the properties of multi-photon and multi-atom bound states in a single-mode waveguide-QED setup, and the associated spectral features are analyzed in detail. In comparison,  references \cite{shi16} and \cite{shi18} discuss the open dynamics of quantum emitters coupled to the bath with a periodic structure, focusing on multi-photon bound states and the construction of a long-range  effective Hamiltonian for quantum emitters. Furthermore, the analytical formulas for both the energy and wave function of bound states has been obtained via variational ans\"atz in the two- and three-excitation subspace. Recently, a thorough study on the collective dynamics of two excited quantum emitters coupled to a one-dimensional structured waveguide has been presented \cite{lanuza}, in which an exact equation for two-excitation bound state is established and solved analytically.  

In light of these advances, previous studies focused mainly on situations in which the bath spectrum exhibits a nontrivial band structure. By this restriction, it is convenient to derive the system dynamics, whether numerical or analytically. However, the excitation dynamics in this situation may be correlated strongly to the bath structure, that make the study too specific to reach the general conclusion.  To tackle this problem, distinct approaches have been proposed to deal with the multiple-excitation dynamics of systems for a general setting of environment, such as matrix-produce-state method \cite{carmele20},  Green's-function method \cite{schneider16, facchi18}, Lehmann-Symanzik-Zimmermann reduction approach \cite{shi09} or Bethe-ansatz technique \cite{shen07}. Although some formulas have been obtained by these ways,  it is still lack of an exactly closed equation, as simple as the single-excitation one.

In this work, we derive the exactly closed equation for the double- or triple-excitation dynamics in a two- or three-emitter system coupled to environment, which is so compact and can be solved accurately. By the equation, the critical condition  for the emergence of  double- and triple-excitation bound state can be determined, which demonstrates the university of multi-excitation bound state in open quantum systems.   As an exemplification, we solve the corresponding equations when quantum emitters  are coupled to Ohmic environment. The calculation shows that  the double- or triple-excitation bound state corresponds to the unique lowest-lying level of quantum emitters with negative energy. In contrast to the single-excitation bound state,  they exhibit robustness against dissipation owing to the  energy gap separating them from the environmental continuum. Furthermore, the critical value of the coupling strength between system and the environment is evaluated.  To verify the derivation, we simulate the real-time dynamics of emitters  using the complex discretization approximation method \cite{cui25}. The simulation agrees well with the derivation.  The paper is organized as follows. In section II, a general description of the quantum emitters and the environment is presented. The section III presents a comprehensive study of the double-excitation dynamics of two quantum emitters coupled to environment. The section IV focuses on the triple-excitation dynamics for three quantum emitters coupled to an environment. Finally, further discussion and conclusion are presented in section V and VI.

\section{The model}
We consider $N$ interacting two-level quantum emitters coupled to a bosonic environment. The
total Hamiltonian is written as
\be\label{h}
H&=&H_s+ \sum_k \omega_k b^{\dagger}_k b_k + \sum_{i, k} g_k^*\sigma^{+}_i b_k + g_k \sigma_i^{-} b^{\dagger}_k, \\
H_s&=&  \sum_{i=1}^{N} \Delta_n n_i +   \lambda \left(\sigma_i^{+}\sigma_{i+1}^{-} + \sigma_i^{-}\sigma_{i+1}^{+} \right)+ U  n_i n_{i+1}, \nonumber
\ee
in which $n_i=\sigma_i^{+}  \sigma_i^{-}$ and the periodic boundary condition $\sigma^{\pm}_{N+1}=\sigma^{\pm}_1$ is imposed. $\sigma_i^{+\left(- \right)}$ represents the raising (lowering) operator for the $i$-th emitter. $\Delta_n=\Delta$ depicts the onsite potential, which is supposed to be isotropic for all emitters.  $\lambda$ characterizes the hopping strength of excitation  between the  nearest neighbored emitters. For clarity, we set $\lambda \equiv 1$ as unit of energy. $U$ quantifies the strength of interaction between two nearest neighbored excited emitters.  For simplicity, both $\Delta$ and $U$ are supposed nonnegative by default in the following study. 

Environment is composed of bosonic modes, differentiated  by symbol  $k$. The coupling coefficient between emitters and environment is given by $g_k$ and its complex conjugate $g_k^*$. Phenomenologically, one can introduce the spectral density function $J(\omega)$ to capture the influence of structureless environment \cite{hp02}, which is defined as
\be \label{J}
J(\omega)=\sum_k \left|g_k\right|^2 \delta\left(\omega -\omega_k \right).
\ee
Thus, the open dynamics of emitters are  determined solely by $J(\omega)$. The rotating-wave approximation is imposed for coupling  such that the number of total excitations $N_e=\sum_{i=1}^{N} \sigma_i^{+}  \sigma_i^{-} + \sum_k b^{\dagger}_k b_k $ is conserved. Thus, the total Hilbert space can be decomposed into a number of subspaces decided by $N_e$. In the calculation below, $\Delta \geq 0$ and $U\geq 0$ are assumed by default.

To determine the bound state, one has to solve the stationary Schr\"odinger equation
\be \label{seqn}
H\ket{\psi_N}=E\ket{\psi_N},
\ee
in which $\ket{\psi_N}$ denotes the total wave function as for $N$ excited emitters. The bound state corresponds to the solution  $E$ lying outside of environment continuum.  For theoretical clarity and definiteness, we focus on the multi-excitation bound state when $N_e=N$, namely the state with  $N$ excitations  localized about $N$ quantum emitters. In this regime, the  bound state is  unique and carries a transparent physical interpretation: it constitutes the unique gapped ground state of an effective Hamiltonian. This property enables a description of the emergence of the multi-excitation bound state from the perspective of dissipative phase transitions. Generalization of the current study  to  the case $N_e<N$ is straightforward. However, bound states  are generally nonunique in this scenario, which substantially complicates the analysis. We therefore omit this case in the present work.  

The remaining discussion focuses on the cases $N_e=2$ and $3$ respectively. Our aim is to derive  the exact equation for the probability amplitude when  $N_e$ emitters are excited simultaneously, the solution of which is the $N_e$-excitation bound state. This amplitude also decides the survival probability of $N_e$ excitations  against dissipation, of which the value can be decided via the Laplace transformation method. Moreover,  we simulate the exact dynamics of the emitters by complex discretization approximation to verify the derivation.

\section{Double-excitation bound state}

In this case, $\ket{\psi_2}$ can be written explicitly as
\be \label{psi2}
\ket{\psi_2} &=& \left( c_0 \sigma_1^+ \sigma_2^+ + \frac{ \sigma_1^+ + \sigma_2^+}{\sqrt{2}} \sum_k c_k b^{\dagger}_k + \sum_{k<l} c_{kl} b^{\dagger}_{k}b^{\dagger}_l +\right. \nonumber  \\
&& \left. \sum_k c_{kk} \frac{ {b^{\dagger}_k}^2 }{\sqrt{2}} + \frac{ \sigma_1^+ - \sigma_2^+}{\sqrt{2}} \sum_{k'} c_{k'} b^{\dagger}_{k'} \right)\ket{0},
\ee
in which $\ket{0}$ denotes the vacuum state. $c_0$ is the probability amplitude for two emitters being excited simultaneously,  of which the finite value signifies the emergence of double-excitation bound state.  $\left(\sigma_1^+ \pm \sigma_2^+ \right)/\sqrt{2}$ characterize the situation that only one of the two emitters is excited. The symmetric form will lead to great simplification in calculation.

Remarkably, it is noted that 
\be
H \left(\frac{ \sigma_1^+ - \sigma_2^+}{\sqrt{2}}\ket{0}\right)=\left(\Delta-\lambda \right) \left(\frac{ \sigma_1^+ - \sigma_2^+}{\sqrt{2}}\ket{0}\right),
\ee
independent of the status of environment. This property is a result that $\left(\sigma_1^+ - \sigma_2^+ \right)/\sqrt{2} \ket{0}$ is a singlet state with zero angular momentum, and thus is immune to dissipation induced by environment. Thus, it is direct
\be
H\left( \frac{ \sigma_1^+ - \sigma_2^+}{\sqrt{2}}b^{\dagger}_{k'}\ket{0}\right)= \left(\Delta-\lambda+\omega_{k'} \right) \left(\frac{ \sigma_1^+ - \sigma_2^+}{\sqrt{2}}b^{\dagger}_{k'}\ket{0}\right). \nonumber
\ee
Namely,  $\left(\sigma_1^+ - \sigma_2^+ \right)/\sqrt{2} b^{\dagger}_{k'}\ket{0}$ is an eigenstate of $H$ with eigenvalue $\Delta-\lambda+\omega_{k'} $. Physically, this state depicts the steady behavior that one excitation is preserved  in emitters, while another is absorbed by environment mode $k'$. Resultantly,  the last term in Eq. \eqref{psi2} is decoupled from the others. Thus, the remaining discussion in this section will focus only on the former four terms in Eq. \eqref{psi2}, which give rise to the double-excitation bound state.

\subsection{Exact equation for double-excitation bound state}

Substituting Eq. \eqref{psi2} into Schr\"odinger equation, one obtains
\be \label{c02}
&\left(2\Delta +U -E \right) c_0  + \sqrt{2}\sum_k g_k^* c_k=0,  \\
&\label{ck2}\left(\Delta +\lambda + \omega_k -E \right) c_k + \sqrt{2}g_k c_0 + 2  g_k^* c_{kk} + \nonumber \\
& \sqrt{2} \sum_{l(\neq k)}  g_l^* c_{kl} =0, \\
&\label{ckk2}\left( 2\omega_k - E \right) c_{kk} + 2 g_k c_k =0, \\
&\label{ckl2}\left(\omega_k + \omega_l - E \right)c_{kl} + \sqrt{2}\left(g_k c_l + g_l c_k \right)=0.
\ee
From Eqs. \eqref{ckk2} and \eqref{ckl2}, we get
\be
c_{kk}&=& \frac{2 g_k}{E - 2 \omega_k} c_k, \\
c_{kl}&=&\frac{\sqrt{2}\left(g_k c_l + g_l c_k \right)}{E - \omega_k - \omega_l}.
\ee
Then, substituting the relations above into Eq. \eqref{ck2} and collecting the items by $c_k$, we have
\be
\left(\Delta +\lambda + \omega_k -E  + 2 \sum_l \frac{\left|g_l \right|^2}{E -\omega_k -\omega_l}\right) c_k  + \nonumber \\
 \sqrt{2}g_k c_0 + 2 g_k \sum_l \frac{g_l^*}{E- \omega_k -\omega_l} c_l =0.
\ee
From the definition of $J(\omega)$ Eq. \eqref{J}, it is direct
\be
\sum_l \frac{\left|g_l \right|^2}{E -\omega_k -\omega_l} = \int_{0 }^{\infty} \text{d}\omega \frac{J(\omega)}{E - \omega_k -\omega}.
\ee
Thus, $c_k$ can be rewritten formally as
\be \label{ck21}
c_k = \frac{-1}{\Delta(\omega_k)} \left(\sqrt{2}g_k c_0 +  2 g_k \sum_l \frac{g_l^*}{E- \omega_k -\omega_l} c_l\right),
\ee
in which  $\Delta(\omega_k)$ is defined as
\be\label{delta2}
\Delta(\omega_k)= \Delta +\lambda + \omega_k -E  + 2 \int_{0 }^{\infty} \text{d}\omega' \tfrac{J(\omega')}{E - \omega_k -\omega'}.
\ee
As illustrated in Appendix \ref{appendix:sbs}, $\Delta(\omega_k)=0$ actually decides a single-excitation bound state with energy $E- \omega_k$. For this reason, the occurrence  of double-excitation bound state is affected significantly by the single-excitation bound state.

Noted that Eq.\eqref{ck21} is a self-consistent equation for $c_k$, $c_k$ can be solved formally as
\be\label{comega2}
c_k &=& -\frac{\sqrt{2}g_k c_0}{\Delta(\omega_k)} c(\omega_k), \\
c(\omega_k)&=&1 - 2\int_{0}^{\infty} \text{d} \omega' \frac{J\left( \omega'\right)}{\Delta\left( \omega'\right) \left(E -\omega_k- \omega'\right)} c\left( \omega'\right).
\ee
Substituting Eq. \eqref{comega2} into Eq. \eqref{c02}, we have
\be
\left(2\Delta +U -E - 2\int_{0}^{\infty} \text{d} \omega \frac{J\left( \omega\right)}{\Delta\left( \omega\right) } c\left( \omega\right) \right) c_0=0.
\ee
For a nonzero $c_0$, one gets directly
\be \label{e2}
2\Delta +U -E - 2\int_{0}^{\infty} \text{d} \omega \frac{J\left( \omega\right)}{\Delta\left( \omega\right) } c\left( \omega\right) =0,
\ee
of which the solution satisfying $E<0$ corresponds to  double-excitation bound state energy.

\subsection{ Determination of $E$}

Eq. \eqref{e2} together with Eq. \eqref{comega2} constitute the central results of this section.  Their solution is, however, not tractable analytically because of the coupled integral and the implicit dependence of $c(\omega)$. To tackle this difficulty, we evaluate $c(\omega)$ and the  integral in Eq. \eqref{e2} using Gauss quadratures (GQ) method.

Different from the equally spaced abscissas in the Newton-Cotes formula, GQ allows the
freedom to choose the location of  abscissas such that the integral is evaluated more efficiently and precisely \cite{wilf}. The central step of GQ is to determine orthogonal polynomials through properly chosen weight function. Thus, the following equality can be obtained
\be \label{gq}
\int_{a}^{b}\text{d}x w(x)f(x)= \sum_{j=1}^{n} w_j f(x_j),
\ee
in which $a$ and $b$ are real, $w(x)$ is the weight function and  $f(x)$ is a polynomials of degree $\leq (2n-1)$.  $x_j$ denotes the zero point of a orthogonal polynomials of degree $n$ and $w_j$ is its weight, both of which can be decided by the recurrence relation for orthogonal  polynomials. As for a series $f(x)$, the integral can be evaluated asymptotically by increasing $n$. 

Regarding Eq. \eqref{e2} and \eqref{comega2}, it is convenient to choose $J(\omega)$ as the weight function. One thus obtains 
\be
\int_{0}^{\infty} \text{d} \omega' \frac{J\left( \omega'\right)}{\Delta\left( \omega'\right) \left(E -\omega_k- \omega'\right)} c\left( \omega'\right) \nonumber \\ \simeq \sum_j w_j \frac{c(\omega_j)}{\Delta\left( \omega_j\right) \left(E- \omega_k - \omega_j \right)}
\ee
Substituting the relation into Eq. \eqref{comega2}, we have
\be
c(\omega_k)=1- 2 \sum_j  \frac{w_j}{\Delta\left( \omega_j\right) \left(E- \omega_k - \omega_j \right)} c(\omega_j).
\ee
This equation can be rewritten in a matrix form as
\be \label{comegaj}
\left(\begin{array}{c}
        c\left(\omega_1\right) \\
        c\left(\omega_2\right) \\
        \vdots \\
        c\left(\omega_n\right)
      \end{array} \right)= \left(\begin{array}{c}
        1 \\
        1 \\
        \vdots \\
        1
      \end{array} \right)- W\cdot  \left(\begin{array}{c}
        c\left(\omega_1\right) \\
        c\left(\omega_2\right) \\
        \vdots \\
        c\left(\omega_n\right)
      \end{array} \right),
\ee
in which the element of matrix $W$ is defined as
\be
W_{kj}=\frac{2 w_j}{\Delta\left( \omega_j\right) \left(E- \omega_k - \omega_j \right)}.
\ee
By this form, $c\left(\omega_j\right)$ can be calculated readily by Cramer rules. Similarly, the integral in Eq. \eqref{e2} can also be approximated as
\be
\int_{0}^{\infty} \text{d} \omega \frac{J\left( \omega\right)}{\Delta\left( \omega\right) } c\left( \omega\right) \simeq \sum_j w_j \frac{c\left( \omega_j \right)}{\Delta\left( \omega_j\right) }.
\ee
Consequently, the double-excitation bound state  can be identified as the $E<0$ satisfying Eq.\eqref{e2}, which should converge to a stable value with increasing $n$. 

Although Eq. \eqref{e2} can be handled numerically by GQ,  a singularity may appear whenever $\Delta\left( \omega_j\right) =0$. This singularity  signals a single-excitation bound state, in  which one excitation becomes localized on an emitter while the other is absorbed by mode $\omega_j$ in environment.  In practice,   the singularity manifests as a sharp divergence of Eq. \eqref{e2} at some particular values  of $E$, readily observed  by plotting  Eq. \eqref{e2} versus  $E$ in the region $E<0$.  Crucially, however, the singularity  is a numerical artifact:  its location is tied to the zeros of the orthogonal polynomials of degree $n$ used in the calculation, and therefore shifts with $n$.  The double-excitation bound state, by contrast,  is a physical solution, independent of the orthogonal polynomials.Thus, its energy  can  be identified unambiguously as the convergent root that  stabilizes upon increasing $n$. Moreover, we have proved in Appendix \ref{appendix:proof} that the double-excitation bound state energy always lies below the singular $E$. By this feature, the double-excitation bound state can be regarded as the ground state of an effective many-body system.

\subsection{Evaluation of $c_0$}\label{sec:c0}
$c_0$ is the probability amplitude  of two excitations being  simultaneously localized on two emitters, a typical feature of the  double-excitation bound state.  To decide its value, one may evaluate the time evolution for initial state with $c_0(0)=1$. Consequently,  the survival probability of initial state is decided absolutely by its overlap to the unique double-excitation bound state, namely $\left| c_0 \right|^2$.

Rather than solving the Schr\"odinger equation $\frac{\partial}{\partial t} \ket{\psi_2(t)} = H  \ket{\psi_2(t)}$ completely, we use the Laplace transformation method to find the long time behavior of $c_0(t)$. The Laplace transformation for  $c_0(t)$ and its derivative to $t$ are written as
\be
\begin{aligned}
c(p)=\int_{0}^{\infty} \text{d}t c(t) e^{- p t},   \\
p c(p) - c(0) =  \int_{0}^{\infty} \text{d}t \frac{\partial c(t)}{\partial t} e^{- p t},
\end{aligned}
\ee
in which $p$ is a complex number with positive real part, and  $c(0)$ denotes the initial condition.  Then,  the equation $\text{i}\partial \ket{\psi_2(t)} /\left( \partial t\right) =H \ket{\psi_2(t)}$  is transformed as 
\be
&\left(2\Delta +U - \text{i}p \right) c_0(p)  + \sqrt{2}\sum_k g_k^* c_k(p)=- \text{i}c_0(0),  \\
&\left(\Delta +\lambda + \omega_k - \text{i}p \right) c_k(p) + \sqrt{2}g_k c_0(p) + 2  g_k^* c_{kk}(p) + \nonumber \\
& \displaystyle \sqrt{2} \sum_{l(\neq k)}  g_l^* c_{kl}(p) =0, \\
&\left( 2\omega_k - \text{i}p \right) c_{kk}(p) - 2 g_k c_k(p) =0, \\
&\left(\omega_k + \omega_l - \text{i}p \right)c_{kl}(p) + \sqrt{2}\left(g_k c_l(p) + g_l c_k(p) \right)=0,
\ee
in which we have chose  $c_0(0)=1$ and $c_{k}(0)=c_{kk}(0)=c_{kl}(0)=0$. If replacing  $\text{i}p$ by $E$, the equations above is same as Eqs. \eqref{c02}-\eqref{ckl2},  except of the term $- \text{i}c_0(0)$. Without loss of generality, we replace $\text{i}p$ by $E$ in the remaining discussion, which is convenient to demonstrate the influence of bound state.

After eliminating the dependence on  $c_{k}(p), c_{kk}(p)$ and $c_{kl}(p)$, we have
\be
c_0(p)= -\frac{\text{i} c_0(0)}{2\Delta +U -E - 2\int_{0}^{\infty} \text{d} \omega \frac{J\left( \omega\right)}{\Delta\left( \omega\right) } c\left( \omega\right)}.
\ee
According to  the inverse Laplace transformation, we get
\be\label{cot2}
c_0(t)&=& \frac{1}{2\pi \text{i}} \int_{s- \text{i}\infty}^{s+ \text{i}\infty} \text{d}p c_0(p)e^{p t}\nonumber \\
&=&  \frac{ 1}{2\pi} \int_{-\text{i}s- \infty}^{-\text{i} s+ \infty}  \text{d} E  \frac{\text{i} c_0(0) e^{- \text{i}E t }}{2\Delta +U -E - 2\int_{0}^{\infty} \text{d} \omega \frac{J\left( \omega\right)}{\Delta\left( \omega\right) } c\left( \omega\right)}.\nonumber \\
\ee
Eq. \eqref{cot2} can be evaluated by residue theorem. Evidently, the real pole of integrand in the above equation is just the solution to Eq. \eqref{e2}, which demonstrates unequivocally influence of double-excitation bound state on the dynamics of emitters.

However,  the complex pole can occur for Eq. \eqref{cot2}, which gives rise to  dissipative evolution  and  becomes ineffective when $t \rightarrow \infty$. Since our focus is the robustness of bound state against dissipation,  Eq. \eqref{cot2}  is evaluated  only for  $t \rightarrow \infty$. Consequently,
one obtains
\be \label{cot22}
c_0(\infty)&=&\lim_{t\rightarrow \infty} c_0(t) \nonumber \\
&=& \lim_{E\rightarrow E_d} \frac{  - c_0(0) e^{- \text{i}E t } \left(E -E_d \right)}{2\Delta +U -E - 2\int_{0}^{\infty} \text{d} \omega \frac{J\left( \omega\right)}{\Delta\left( \omega\right) } c\left( \omega\right)},
\ee
where $E_d$ denotes the real solution to Eq. \eqref{e2}, namely the double-excitation bound state energy. To obtain Eq. \eqref{cot22}, we has utilized the fact that  $E_d$ is a pole of degree one. Thus, Eq. \eqref{cot22} can be evaluated readily by L'H\^{o}pital's rule.

\subsection{Exemplification: Ohmic bath}

\begin{table}
  \centering
  \begin{tabular}{c|c|c}
    \hline \hline
   $n$  & $E_d$ & $\left|c_0\left(\infty \right) \right|^2$ \\
     \hline
    $50$ & -0.75045 &  0.49953\\
    $100$ & -0.75209 & 0.49221\\
    $150$ & -0.75221 & 0.49167\\
    $200$ & -0.75223 & 0.49122\\
    $250$ & -0.7522325& 0.491183  \\
    $300$ & -0.7522332 &  0.491174\\
     $400$ & -0.7522335 &  0.4911702\\
    \hline  \hline
  \end{tabular}
  \caption{The double-excitation bound state energy $E_d$ and survival probability $\left|c_0\left(\infty \right) \right|^2$ determined by solving Eq. \eqref{e2} and Eq. \eqref{cot22} respectively, versus with the degree $n$ for established orthogonal polynomials by Ohmic spectral density function Eq. \eqref{ohm}. For this calculation, we choose $\Delta=0.5, U=0, \eta=0.1$ and $\omega_c=10$}\label{table:nscaling}
\end{table}

\begin{figure}
\center
\includegraphics[width=6cm]{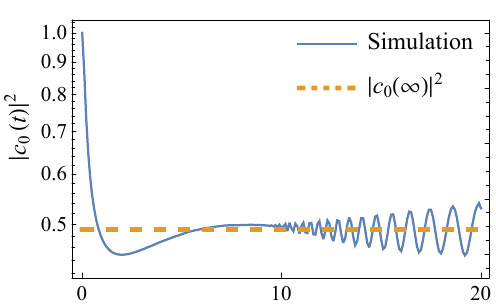}
\caption{(Color online) A comparative plot for $\left|c_0\left(\infty \right) \right|^2=0.491174$ (dashed line) and the evolution $\left|c_0\left(t\right) \right|^2$ (solid line), simulated by CDA. As for CDA, we have chosen the radius for the integral path in complex plan as $3\omega_c$ and the degree of orthogonal polynomials is 150. The other parameters are the same as those in Table \ref{table:nscaling}. The obvious oscillation of $\left|c_0\left(t\right) \right|^2$ stems from the recurrence induced by finite dimensionality of effective Hamiltonian.   }
\label{fig:simulation}
\end{figure}

As an exemplification, we consider Ohmic bath to demonstrate the emergency of double-excitation bound state.  The Ohmic spectral density function is defined as
\be \label{ohm}
J(\omega)= \eta \omega e^{-\omega/\omega_c},
\ee
where $\omega_c$ is the cut-off frequency, and $\eta>0$ is the coupling strength. By choosing $J(\omega)$ as weight function, the established orthogonal polynomials corresponds to  Laguerre polynomials.

We first demonstrate the $n$-scaling behavior for the double-excitation bound state energy $E_d$ and survival probability $\left|c_0\left(\infty \right) \right|^2$  respectively. As shown in Table \ref{table:nscaling}, both $E_d$ and $\left|c_0\left(\infty \right) \right|^2$ converge to a stable value by increasing  $n$. Furthermore, we also simulate the exact time evolution of $\left|c_0\left(t\right) \right|^2$ by complex discretization approximation (CDA) method (see Ref. \cite{cui25} for a comprehensive introduction).  As demonstrated by Fig.\eqref{fig:simulation}, $\left|c_0\left(t\right) \right|^2$ illustrates good compliance with  the predication from Eq. \eqref{cot22}. The fluctuation of $\left|c_0\left(t\right) \right|^2$ in Fig.\eqref{fig:simulation} stems from the recurrence induced by finite dimensionality of effective Hamiltonian. 

\begin{figure}
\center
\includegraphics[width=5cm]{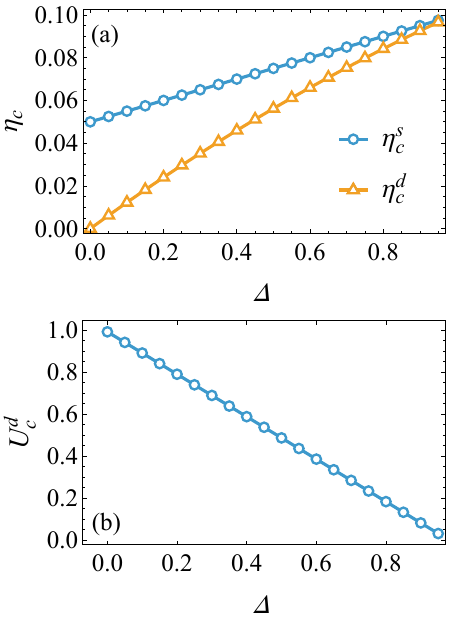}
\caption{(Color online) (a) A comparative plotting of $\eta_c^s$ (open triangle), $\eta_c^d$ for $U=0$. To determine $\eta_c^d$ , we solve Eq. \eqref{e2} by setting $E=0$ with $n=200$.  (b) Plots for $U_c^d$.  $U_c^d$ is determined through calculating Eq. \eqref{e2} directly  by setting $U=E=0$ and $\eta= \eta_c^s$.   For all plots, $\omega_c=10$ are chosen. }
\label{fig:etac}
\end{figure}

\begin{figure}
\center
\includegraphics[width=\columnwidth]{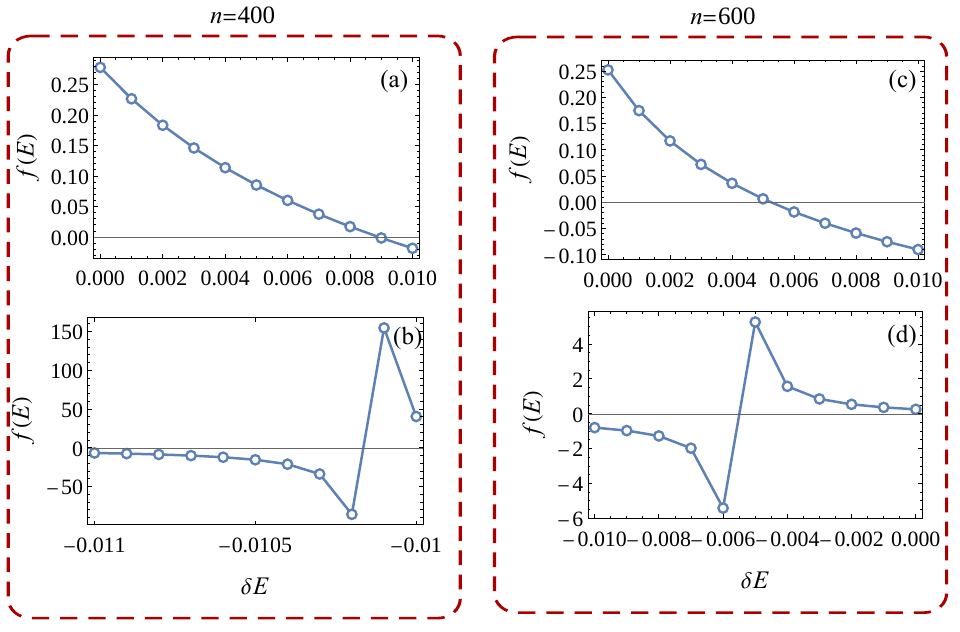}
\caption{(Color online) The diagrammatic sketch for the emergence of double-excitation bound state and its dependence on the degree $n$ of orthogonal polynomials. For the plots, we have chosen $\Delta=1, U=0, \omega_c=10$ and $\eta=0.11$, larger than $\eta_c^s=0.1$ by Eq. \eqref{setac}. $f(E)$ is defined as the left hand side of Eq. \eqref{e2}. $\delta E= E- E_k$, $E_k$ being the solution to $\Delta(\omega_k)=0$ (Eq. \eqref{delta2}) with a definite $\omega_k$, which is varying with $n$.  $E_k\sim -0.01458$ for $n=400$, while  $E_k\sim -0.04504$   for $n=600$. }
\label{fig:f(e)}
\end{figure}

A critical coupling $\eta_c^d$ can be obtained by solving Eq. \eqref{e2} with $E=0$,  which marks the emergence of double-excitation bound state for $\eta> \eta_c$. Direct calculation is, however,  complicated by the singularity associated with $\Delta(\omega)=0$. The explicit evaluation reveals two distinct behaviors  for $\eta_c^d$ , depending on  $\Delta<1$ or $>1$. As shown in Fig.\ref{fig:etac} (a),  $\eta_c^d$ can be uniquely determined  for  $\Delta<1$ and $U=0$, which lies below the single-excitation threshold $\eta_c^s$ defined in Eq. \ref{setac}. The double-excitation bound state therefore emerges before its single-excitation counterpart as $\eta$ is increased. Increasing $U (>0)$ leads to a larger $\eta_c^d$. We may accordingly define a critical interaction $U^d_c$ as the largest $U$ for which $\eta_c^d \leq \eta_c^s$ is hold.  As illustrated in Fig.\ref{fig:etac} (b), $U^d_c$ tends to be zero when $\Delta \rightarrow 1$ from below, indicating a qualitative change at $\Delta=1$. For $U>U^d_c$, $\eta_c^d$ would be large than $\eta_c^s$ such that the singularity induced by $\Delta(\omega)=0$ makes the calculation obscured. Fig. \ref{fig:f(e)} shows that  the singularity is highly sensitive to $n$: not only the zero point of Eq. \eqref{e2} can  be shifted appreciably even if $n$ is large ( panel (a) and (c)), but also the evaluation may exhibit strong numerical fluctuations nearsingular point, as shown by panel (b) and (d). 

\begin{figure}
\center
\includegraphics[width=8cm]{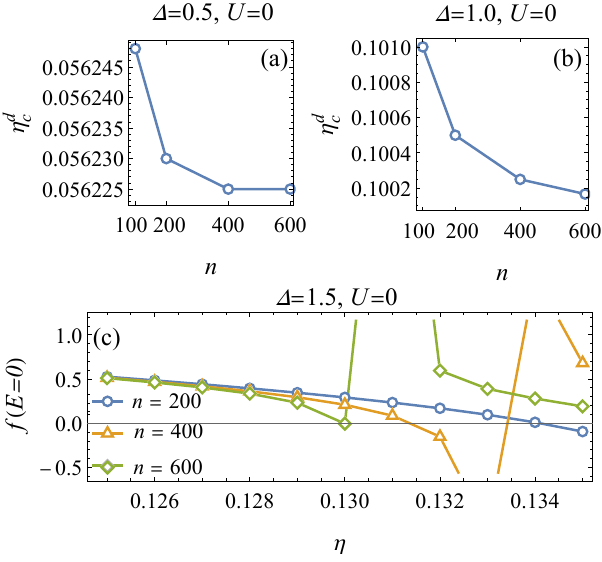}
\caption{(Color online) The $n$ scaling behavior of determined  $\eta_c^d$  for $\Delta=0.5$ (a) and $\Delta=1.0$ (b). However for $\Delta=1.5$, $f(E)$ is calculated for $\eta > \eta_c^s=0.15$ because of the absence of  $\eta_c^d$. For all plots, $E=U=0$ and $\omega_c=10$ are chosen.  }
\label{fig:vlone}
\end{figure}

For $\Delta \geq 1$, by contrast, the determination of $\eta_c^d$ becomes problematic.   Fig. \ref{fig:vlone} (a), (b) and (c) compare the convergence of $\eta_c^d$  for $\Delta=0.5$, $1$ and $1.5$. Given the rapid convergence when $\Delta=0.5$, the convergence at both $\Delta=1$ and  $\Delta=1.5$ is so slow that $\eta_c^d$ cannot be extracted reliably. Particularly, the evaluation illustrates strong numerical fluctuation as shown by panel (c).  
This obscurity can be traced to the upward trend of $\eta_c^d$ with $\Delta$ [Fig.~\ref{fig:etac}(a)]: once $\eta_c^d\geq\eta_c^s$, the double-excitation threshold is pushed into the very region where the $\Delta(\omega_j)=0$ singularity dominates, and the physical root is no longer separable from the $n$-dependent spurious feature.

\section{triple-excitation bound state}
For $N=3$, the eigenstate for  Eq. \eqref{h} can be written as
\be\label{psi3}
\ket{\psi_3}&=&c_0\ket{3}\otimes \ket{0}  \nonumber \\ &+& \ket{2,1}\otimes \sum_k c_{1,k}b^{\dagger}_k \ket{0}+ \ket{2,2}\otimes \sum_k c_{2,k}b^{\dagger}_k \ket{0}  \nonumber \\ &+& \ket{2,3}\otimes \sum_k c_{3,k}b^{\dagger}_k \ket{0}
+  \ket{1,1}\otimes \sum_{k\leq l} c_{k,l}b^{\dagger}_k b^{\dagger}_l \ket{0}  \nonumber \\
&+&  \ket{1,2}\otimes \sum_{k\leq l} c_{k,l}'b^{\dagger}_k b^{\dagger}_l \ket{0} +  \ket{1,3}\otimes \sum_{k\leq l} c_{k,l}''b^{\dagger}_k b^{\dagger}_l \ket{0} \nonumber \\
&+& \ket{0} \sum_k c_{kkk} \frac{\left(b^{\dagger}_k\right)^3}{\sqrt{3!}}\ket{0}+ \ket{0}\sum_{k\neq l} c_{kll}  \frac{\left(b^{\dagger}_k\right)^2}{\sqrt{2}} b_l\ket{0} \nonumber \\
&+& \ket{0}\sum_{k<l <p} c_{klp} b^{\dagger}_k b^{\dagger}_l b^{\dagger}_p \ket{0},
\ee
where the cyclic symmetry basis states for three emitters are defined as
\be
\ket{3}&=&\sigma^{+}_1 \sigma^{+}_2 \sigma^{+}_3\ket{0},\nonumber \\
\ket{2,1}&=& \frac{1}{\sqrt{3}}\left(\sigma^{+}_1 \sigma^{+}_2 + \sigma^{+}_3 \sigma^{+}_3 + \sigma^{+}_2 \sigma^{+}_3 \right)\ket{0},\nonumber \\
\ket{2,2}&=& \frac{1}{\sqrt{3}}\left(\sigma^{+}_1 \sigma^{+}_2 + e^{\text{i}\frac{2}{3}\pi}\sigma^{+}_1 \sigma^{+}_3 +  e^{-\text{i}\frac{2}{3}\pi}\sigma^{+}_2 \sigma^{+}_3 \right)\ket{0}, \nonumber \\
\ket{2,3}&=& \frac{1}{\sqrt{3}}\left(\sigma^{+}_1 \sigma^{+}_2 + e^{-\text{i}\frac{2}{3}\pi}\sigma^{+}_1 \sigma^{+}_3 +  e^{\text{i}\frac{2}{3}\pi}\sigma^{+}_2 \sigma^{+}_3 \right)\ket{0}, \nonumber \\
\ket{1,1}&=& \frac{1}{\sqrt{3}}\left(\sigma^{+}_1 + \sigma^{+}_2 + \sigma^{+}_3 \right)\ket{0}, \nonumber \\
\ket{1,2}&=& \frac{1}{\sqrt{3}}\left(\sigma^{+}_1 + e^{\text{i}\frac{2}{3}\pi}\sigma^{+}_2  +  e^{-\text{i}\frac{2}{3}\pi} \sigma^{+}_3 \right)\ket{0},\nonumber \\
\ket{1,3}&=& \frac{1}{\sqrt{3}}\left(\sigma^{+}_1 + e^{-\text{i}\frac{2}{3}\pi}\sigma^{+}_2  +  e^{\text{i}\frac{2}{3}\pi} \sigma^{+}_3 \right)\ket{0}.
\ee
The cyclic state can simplify  the calculation substantially. Under the cyclic exchanging the subscript  $1\rightarrow 2 \rightarrow 3 \rightarrow 1$, states $\ket{3}, \ket{2,1}$ and $\ket{1,1}$  are invariant. In contrast,  $\ket{2, 2}$ and $\ket{1, 2}$  acquire a phase factor $e^{- \text{i}\frac{2\pi}{3}}$, while $e^{ \text{i}\frac{2\pi}{3}}$ for $\ket{2, 3}$ and $\ket{1, 3}$.  The  Hilbert space therefore can be decomposed into three  symmetry sectors labeled by the phase factores $\{1,e^{\pm 2\pi\mathrm{i}/3}\}$. Since the triple-excitation bound state has overlap only with $\ket{3}$, it lies entirely in the invariant sector, and the eigenvalue problem reduces to that single subspace. 

By time-independent Schr\"odinger equation, we obtain a system of equation
\be \label{cklp}
&\left[3\left(\Delta + U \right)- E\right]c_0 + \sqrt{3}\sum_k g_k^{*} c_{1,k}=0, \nonumber \\
&\left[2\left(\Delta + \lambda \right)+U + \omega_k- E\right]c_{1,k} + \sqrt{3}g_k c_0 + 2\sqrt{2} g_k^{*} c_{k,k} \nonumber \\  & + 2\sum_{l\left(\neq k \right)} g_l^{*} c_{k,l} =0, \nonumber \\
& \left(\Delta + 2\lambda +\omega_k +\omega_l -E  \right)c_{k, l} + 2\left(g_l c_{1, k} + g_k c_{1,l} \right), \nonumber \\
& \sqrt{6}\left(g_k^{*} c_{kkl} + g_l^{*} c_{llk} \right) + \sqrt{3} \sum_{p\left(\neq k, l \& k<l \right) }  g_p^{*} c_{klp}=0, \nonumber \\
 &  \left(\Delta + 2\lambda +2 \omega_k  -E  \right)c_{k, k} + 2\sqrt{2} g_k c_{1,k}+ 3 g_k^{*} c_{kkk} \nonumber \\
& + \sqrt{3} \left(\sum_{l\left(> k \right)}  g_k^{*} c_{kkl} + \sum_{p\left(< k \right)}  g_p^{*} c_{pkk}   \right) =0, \nonumber \\
&  \left(3 \omega_k  -E  \right)c_{kkk} + 3 g_k c_{k, k}=0, \nonumber \\
&   \left(2 \omega_k +\omega_l  -E  \right)c_{kkl}+\sqrt{6}g_k c_{k,l} + \sqrt{3} g_l c_{kk}=0,\nonumber \\
&  \left( \omega_k +\omega_l + \omega_p -E  \right)c_{klp} + \nonumber \\
& \sqrt{3}\left(g_p c_{k, l} + g_l c_{k, p} + g_k c_{l, p}\right) =0 .
\ee
Eliminating the coefficients  $c_{kkk}, c_{kkl}, c_{klp}$, $c_{k,l} $ and $c_{1,k}$, one obtains an iterative equation for $E$  
\be \label{e3}
3\left(\Delta+U \right)- E + 3 \int_{0}^{\infty} \text{d}\omega_k J\left(\omega_k \right) \overline{c}_{1, k}=0, 
\ee 
in which
\begin{equation} \label{e33}
\begin{aligned}
\overline{c}_{1, k} &=& -\frac{1}{\Delta_k } \left[1 +2 \int_{0}^{\infty} \text{d}\omega_p J\left(\omega_p \right) \overline{c}_{k,p} \right], \\
 \overline{c}_{k,l} &=& -\frac{1}{\Delta_{k,l}} \left[ \overline{c}_{1,k} + \overline{c}_{1,l} +  3 \int_{0}^{\infty} \text{d}\omega_p \times \right.   \\
 && \left.   \frac{J\left(\omega_p \right) }{E- \omega_k -\omega_l  -\omega_p}\left( \overline{c}_{k, p} + \overline{c}_{l, p}\right) \right].
 \end{aligned}
\end{equation}
The explicit definition for $\overline{c}_{1, k}, \overline{c}_{k, l}$ can be found in Appendix \ref{appendix:cklp}. $\Delta_{k, l}$ and $\Delta_{k}$ is defined by Eq. \eqref{dkl} and \eqref{dk}. 

Clearly, a negative solution to Eq. \eqref{e3} gives triple-excitation bound state energy, which can be determined numerically by Gauss quadrature. However, the singularity can appear whenever $\Delta_k=0$ or $\Delta_{k, l}=0$, as demonstrated in Eq. \eqref{e33}. $\Delta_k=0$ yields a  solution with $E>0$ inside the environmental continuum, and thus  has no influence on the bound state. However, $\Delta_{k, l}=0$ signifies a single-excitation bound state with energy $E- \omega_k-\omega_l$, as demonstrated by Eq. \eqref{E3}. Similar as the double-excitation case, it produces  $n$-dependent structures that obscure the physical root.  Although so, the triple-excitation bound state can still be determined accurately by tracking the root that converges as $n$ is increased. Furthermore, we have proved in Appendix \ref{appendix:proof}2 that it can always be found below the single-excitation threshold. 

\begin{figure}
\center
\includegraphics[width=6cm]{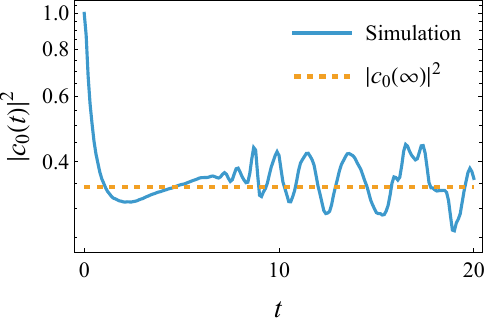}
\caption{(Color online) A comparative plot for $\left|c_0\left(\infty \right) \right|^2=0.34205$ (dashed line) and the evolution $\left|c_0\left(t\right) \right|^2$ (solid line), simulated by CDA. As for CDA, we have chosen the radius for the integral path in complex plan as $\omega_c$ and the degree of orthogonal polynomials is 50. The other parameters are the same as those in Table \ref{table:nscaling3}. The obvious oscillation of $\left|c_0\left(t\right) \right|^2$ stems from the recurrence induced by finite dimensionality of effective Hamiltonian.   }
\label{fig:simulation3}
\end{figure}

As an example, we evaluate the triple-excitation bound state and corresponding $\left|c_0\left(\infty \right) \right|^2$ for $\Delta=0.5, \eta=0.1$ and $U=0$, as shown in Table \ref{table:nscaling3}. Limited by our computational ability, the calculation is carried out up to $n=150$, and thus the numerical accuracy $10^{-3}$ can be reached. Furthermore, to verify the derivation, we also utilize CDA method to simulate the exact dynamics of three emitters initially all excited. As shown in Fig. \ref{fig:simulation3}, the simulation  agrees with $\left|c_0\left(\infty\right) \right|^2$ extracted from Eq.\eqref{e3}, although  $\left|c_0\left(t\right) \right|^2$ oscillates strongly at large $t$.

\begin{table}
  \centering
  \begin{tabular}{c|c|c}
    \hline \hline
   $n$  & $E_t$ & $\left|c_0\left(\infty \right) \right|^2$ \\
     \hline
    $50$ & -1.07746 &  0.34948\\
    $100$ & -1.07964 & 0.34273\\
    $150$ & -1.07978 & 0.34205\\
    \hline  \hline
  \end{tabular}
  \caption{The triple-excitation bound state level $E_t$ and survival probability $\left|c_0\left(\infty \right) \right|^2$ determined by solving Eq. \eqref{e3} versus with the degree $n$ of orthogonal polynomials established from Ohmic spectral density function Eq. \eqref{ohm}. For this calculation, we choose $\Delta=0.5, U=0, \eta=0.1$ and $\omega_c=10$}\label{table:nscaling3}
\end{table}

\begin{figure}
\center
\includegraphics[width=5cm]{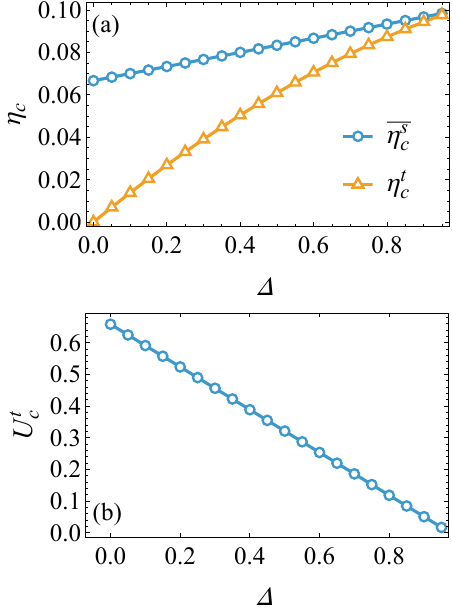}
\caption{(Color online) (a) A comparative plotting of $\overline{\eta_c^s}$ (open circle), $\eta_c^t$ for $U=0$. To determine $\eta_c^t$ , we solve Eq. \eqref{e3} by setting $E=0$ with $n=150$.  (b) Plots for $U_c^t$.  $U_c^t$ is determined through calculating Eq. \eqref{e3} directly with $E=0$ and $\eta= \overline{\eta_c^s}$.  For all plots, $\omega_c=10$ are chosen. }
\label{fig:etac3}
\end{figure}

The critical coupling  $\eta_c^t$ for the emergency of triple-excitation bound state can be determined by solving Eqs.\eqref{e3} for $E=0$. In Fig. \ref{fig:etac3}(a),  both  $\eta_c^t$ and the single-excitation threshold $\overline{\eta_c^s}$ defined by Eq. \eqref{tetac} are plotted for $\Delta<1$. By increasing $\Delta$, $\eta_c^t$  approaches $\overline{\eta_c^s}$ from below, and they two coincide  at $\Delta=1$. For $\Delta \geq 1$, the numerical evaluation is obscured because of the occurrence of single-excitation bound state, and thus  $\eta_c^t$  cannot be evaluated reliably in this case. This is precisely the scenario encountered in the two-emitter case.  Similarly, increasing $U>0$ shifts  $\eta_c^t$  upwards. Fig. \ref{fig:etac3}(b) plots $U_c^t$, the maximal $U$ for which $\eta_c^t \leq \overline{\eta_c^s}$ holds. It is found that  $U_c^t$ tends to be zero with $\Delta\rightarrow 1$.

\section{Discussion }\label{sec:discussion}

We have demonstrated the occurrence of double- or triple-excitation bound state, and its impact on the emitter dynamics. Given these two bound states differ in the number of excitations, they share some common features, which are worthy of further discussion.

\emph{Properties of $\eta_c^d$ or $\eta_c^t$ for $\Delta/\lambda<1$}. It is noted that both  $\eta_c^d$  and $\eta_c^t$  lies below their single-excitation counterparts, and in fact tend to be zero when $\Delta\rightarrow 0$. This feature implies that the many noninteracting excitations would be preserved more easily than the single one.  This reason is  energetic: to form a stable bound cluster, the constituent excitations have to lower their energy below the continuum edge, and the inter-excitation correlations supply the requisite extra lowering. For concreteness,  we evaluate the single-excitation bound state for two or three emitters case respectively when $\Delta=0.5, U=0, \eta=0.1$ and $\omega_c=10$, for which the corresponding double- or triple-excitation bound state have been decided in Table \ref{table:nscaling} or \ref{table:nscaling3}.  For two emitters, Eq. \eqref{sbs2} gives the  single-excitation bound state energy $E_s=-0.31196$. Together with the double-excitation energy $E_d=-0.75223$ from Table~\ref{table:nscaling}, this yields a binding energy
\be
E_{\text{bind}} = 2 E_s - E_d= 2\times (-0.31196)+0.75223= 0.12831. \nonumber
\ee
For three emitters,  Eq. \eqref{sbs3} gives the  single-excitation bound state energy $E_s=-0.25522$. Together with the double-excitation energy $E_d=-1.07978$ from Table~\ref{table:nscaling3}, one obtains
\be 
E_{\text{bind}} = 3 E_s - E_t=  3\times (-0.25522)+1.07978=0.31412. \nonumber
\ee

The coupling  provides the channel of releasing the excess energy $E_{\text{bind}}$ into environment. To guarantee the this process irreversible, a weak coupling is favored, so that the emitted energy can not flow back to emitters.  By contrast, preserving  excitation  around emitters require the coupling strong enough such that the exchanging of energy between emitters and environment can become interfered destructively.  Bound-state formation is therefore optimal only at an intermediate coupling, where these two requirements are balanced.

\emph{Effect of  $U$}. Increasing $U (>0)$ raises the bound state energy. The binding energy $E_{\text{bind}}$ therefore shrinks and eventually turns negative. Consequently, the requirement of releasing superfluous energy is relaxed or even reversed, that gives rise to a larger coupling strength. Accordingly, we denote $U_c^d$ or $U_c^t$ respectively as the value of interaction, at which $\eta_c^d$ or $\eta_c^t$  rise to meet the single-particle threshold: for $U$ beyond these valued, the double- or triple-excitation bound state is no longer identified reliable because of the singularity induced by the single-excitation bound state. 

\emph{$\Delta \geq 1$}. In this situation, the convergence of the quadrature becomes very slow, as shown in Figs. \ref{fig:vlone}. Worse still, the root $\eta$ to Eq. \eqref{e2} at $E=0$ is no longer unique: the curve develops two crossings, as shown in Figs. \ref{fig:vlone}(c). $\eta_c^d$ is therefore not extractable in this regime.  The triple-excitation sector behaves the same way.  

In another point, extrapolating the plotted trends in $\eta_c^ {d(t)}$ and $U_c^{d (t)}$ shown Fig.\ref{fig:etac} and \ref{fig:etac3}, we conjecture  that $\eta_c^ {d(t)}$ be larger than its single-excitation counterpart. Unfortunately, we have no satisfactory explanation for this anomalous behavior. Formally, it can be attributed to  the occurrence of single-excitation bound state, which intrigues  the divergence of  Eq. \eqref{e2} or \eqref{e3}. To settle this problem, a natural way is to  bypass the integral equation altogether and obtain the spectrum directly, for example by exact diagonalization, such as the discretization approximation \cite{chin10} or CDA \cite{cui25}. Then,  the multiple-excitation bound state appears as the ground state. However, this approach would be consumptive and exhaustive, and is out of our computational resource. 

It should be stressed, however,  that though the absence of critical $\eta$ for $\Delta \geq 1$, the double- or triple-excitation bound state can still be determined accurately for any $\Delta$, even if the coupling is above the single-excitation threshold. The main step is to identify the zero point for Eq. \eqref{e2} or \eqref{e3}, which converges into a stable value as increasing $n$.

\section{Conclusion}

In conclusion, we have studied the double- or triple-excitation dynamics respectively in a system of  two  or three quantum emitters coupled to  environment.  In both cases, the exact equation can be  established for the bound state, as illustrated by Eq. \eqref{e2} or Eq. \eqref{e3}. Consequently, the bound state energy and the associated probability amplitude $c_0$ can be decided accurately by employing the Gauss quadrature method.  To verify our derivation, we  utilize CDA method to simulate the exact dynamics for excited emitters coupled to Ohmic environment. The two approaches yield the consistent results. Moreover, the critical condition for the emergence of bound state can be determined uniquely for $\Delta<1$. This picture indicates that the occurrence of double- or triple-excitation bound state can be understood as a dissipative  phase transition. However, for $\Delta \geq 1$, the evaluation becomes obscured  owing to the singularity induced by occurrence of single-excitation bound state, which makes the evaluation obscure.

We have demonstrated that the occurrence of multiple-excitation bound state is universal in open quantum systems, irrespective of  whether environment is structured or structureless. Our finding has significant physical implications. First, our study generalizes the exactly solvable situation into the multi-excitation realm, and thus provides an exact viewpoint into the dynamics of open interacting quantum systems. Secondly, the multiple-excitation bound state  provides a way to manipulate the interaction among quantum emitters, which is crucial for quantum computation.  Thirdly, the  multiple-excitation bound state provides an alternative way to realise storage of energy. In light of  the advances in quantum batteries \cite{QB},  it is possible to stabilize quantum batteries against dissipation by mediating the coupling.  Finally, the finding that many excitations would be preserved  more easily than a single one,  implies that one could control the transmission of single excitation by multi-excitation bound state, which seems provide another way to realize photon blockade \cite{birnbaum05}. 

\acknowledgments
H. T. Cui acknowledges the support of Provincial Natural Science Foundation (ZR2021MA036). Y. A. Yan acknowledges the support of National Natural Science Foundation of China (22450002).   H. Z. Shen acknowledges the supports of Science and Technology Development Plan Project of Jilin Province (20250102007JC) and National Natural Science Foundation of China (12274064).

\appendix

\section{The bound state in the single-excitation subspace }

\subsection{$N=2$}\label{appendix:sbs}
For two emitters, the single-excitation bound state can be written formally as
\be
\ket{\psi_1}=\left(c_1\sigma^+_1 + c_2 \sigma_2^+ + \sum_k c_k b^{\dagger}_k \right)\ket{0}.
\ee
Substituting $\ket{\psi_1}$ into Schr\"odinger equation, one gets
\be
\Delta c_1 + \lambda c_2 + \sum_k g_k^* c_k &=& E c_1, \\
\Delta c_2 + \lambda c_1 + \sum_k g_k^* c_k &=& E c_2, \\
\omega_k c_k + \left(c_1 +c_2 \right)g_k &=& E c_k.
\ee
Eliminating $c_k$ by the third equation, one has
\be
\Delta c_1 + \lambda c_2 + \left(c_1 +c_2 \right)\int_{0}^{\infty}\text{d}\omega \frac{J(\omega)}{E-\omega}&=& E c_1, \\
\Delta c_2 + \lambda c_1 + \left(c_1 +c_2 \right) \int_{0}^{\infty}\text{d}\omega \frac{J(\omega)}{E-\omega} &=& E c_2,
\ee
which is a linear system of equations for $c_1$ and $c_2$. It is ready to obtain two eigenvalues,
\be
& E_1= \Delta- \lambda, \\
& \label{sbs2}\displaystyle \Delta+\lambda- E_2 + 2 \int_{0}^{\infty}\text{d}\omega \frac{J(\omega)}{E_2-\omega}=0.
\ee
Evidently, level $E_1$ is decoupled  to environment, and thus is decoherence-free state. The corresponding eigenstate $\ket{E_1}= \left(\sigma_1 -\sigma_2 \right)/\sqrt{2} \ket{0}$ denotes a singlet, which is physical reason for its immunity to decoherence. 

$E_2$ denotes the single-excitation bound state energy, which can be figured out by solving Eq. \eqref{sbs2} in the region $E_2 \leq 0$. Noted the left hand side of Eq. \eqref{sbs2} decreases monotonically with increase of  $E_2$,  a critical coupling can be obtained by setting $E_2=0$ in Eq. \eqref{sbs2}. As an example, for the Ohmic spectral density function, a critical coupling can be determined, 
\be \label{setac}
\eta^s_c= \frac{\Delta+\lambda}{2 \omega_c},
\ee
in which the superscript $s$ denotes the single-excitation. Thus, the single-excitation bound state occures only if $\eta > \eta^s_c$.

\subsection{$N=3$}
For three emitters, the single-excitation bound state can be expressed  as
\be
\ket{\overline{\psi_1}}= c_1 \ket{1,1}+c_2 \ket{1,2} + c_3 \ket{1,3} +\sum_k c_k b^{\dagger}_k \ket{0}.
\ee
Substituting $\ket{\overline{\psi_1}}$ into Schr\"odinger equation, one gets
\be
\begin{aligned}
\left( \Delta + 2\lambda - E \right) c_1 + \sqrt{3}\sum_k g^{*}_k c_k &=& 0,  \\
\left(\Delta -\lambda-E  \right) c_2 &=& 0,  \\
 \left(\Delta -\lambda -E \right) c_3 &=& 0,  \\
 \left( \omega_k -E \right) c_k + \sqrt{3}g_k c_1 &=& 0.
\end{aligned}
\ee
Eliminating  $c_k$, we get
\be \label{sbs3}
\left( \Delta + 2\lambda - E + 3\int_{0}^{\infty}\text{d}\omega \frac{J(\omega)}{E-\omega} \right) c_1=0.
\ee
Thus, the single-excitation bound state energy labelled by $E_3$ in this case can be determined by solving the equation
\be \label{E3}
\Delta + 2\lambda - E_3 + 3\int_{0}^{\infty}\text{d}\omega \frac{J(\omega)}{E_3-\omega} =0.
\ee
The level $\Delta -\lambda$ is decoherence-free, and is doubly degenerate, which is protected by nontrivial geometry embedded in $\ket{1, 2}$ and $\ket{1, 3}$.

The critical coupling can be identified by setting $E_3=0$ in Eq. \eqref{E3}. For Ohmic spectral  density function, one gets 
\be \label{tetac}
\overline{\eta_c^s}=\frac{\Delta + 2\lambda}{3\omega_c},
\ee
beyond which the single-excitation bound state can occur.  

\section{Proof that the energy of double- or triple- excitation bound state is smaller than the single-excitation threshold}\label{appendix:proof}

\subsection{$N=2$}
By Eq. \eqref{e2}, we first prove  $\int_{0}^{\infty} \text{d} \omega \frac{J\left( \omega\right)}{\Delta\left( \omega\right) } c\left( \omega\right)>0$ when $E<0$. From Eq. \eqref{delta2},  $\Delta(\omega)$ decrease monotonically by increasing $E$. Denoting  $E_0(<0) $ as the minimal solution to $\Delta(\omega)=0$, we can find
\be
\Delta(\omega)
\begin{cases}
  >0, & \mbox{if } E<E_0  \\
  \text{undetermined}, & \mbox{if }  E>E_0. 
\end{cases}
\ee
Notably, $E_0$ is also solution to Eq. \eqref{sbs2}, and denotes the single-excitation bound state. 

$c\left(\omega\right)$ is related to  $\Delta(\omega)$. Rewriting  $c\left(\omega\right)$ as iteration
\be
c\left(\omega\right)&=& 1- 2\int_{0}^{\infty} \text{d} \omega' \tfrac{J\left( \omega'\right)}{\Delta\left( \omega'\right) \left(E -\omega- \omega'\right)} +  4\times \nonumber \\
&& \int_{0}^{\infty} \text{d} \omega' \tfrac{J\left( \omega'\right)}{\Delta\left( \omega'\right) \left(E -\omega - \omega'\right)} \int_{0}^{\infty} \text{d} \omega'' \tfrac{J\left( \omega''\right)}{\Delta\left( \omega''\right) \left(E -\omega'- \omega''\right)} + \cdots \nonumber,
\ee
we have $c\left(\omega\right)>0$ when $E<E_0 $ since $\Delta(\omega)>0$ in this case. Consequently, we obtain
\be
\int_{0}^{\infty} \text{d} \omega \frac{J\left( \omega\right)}{\Delta\left( \omega\right) } c\left( \omega\right)>0, \mbox{if }  E<E_0<0.
\ee
Thus, with $\Delta>0$ and $U>0$, the solution to Eq. \eqref{e2} may be found in the region $E<E_0<0$. On the contrary, $c\left(\omega\right)$ is not guaranteed to be positive when $E>E_0$, and thus the solution to Eq. \eqref{e2} may not exist.  Consequently, the double-excitation bound state can occur sufficiently when $E<E_0 $.

For $E>E_0$, the solution to Eq. \eqref{delta2} is $E_0 -\omega_k$, and thus is involved with the environmental mode $\omega_k$. Resultantly, the calculation becomes sensitive to  the degree $n$ of polynomials, and is not general in physics. 

\subsection{$N=3$}
By Eq. \eqref{e33}, $\overline{c}_{k,l}$ can rewritten as 
\be 
\overline{c}_{k,l}= \frac{1}{\Delta_{k,l}} \left[ \frac{1}{\Delta_k} +  \frac{1}{\Delta_l} + \int_{0}^{\infty}\text{d}\omega_p \left(\frac{2}{\Delta_k} - \frac{3}{E_{klp}} \right) \overline{c}_{k, p} \right. \nonumber \\
 + \left. \int_{0}^{\infty}\text{d}\omega_p \left(\frac{2}{\Delta_l} - \frac{3}{E_{klp}} \right) \overline{c}_{l, p} \right], \nonumber \\
\ee
in which $E_{klp}=E - \omega_k -\omega_l -\omega_p$. The definition of $\Delta_k$ and $\Delta_{k,l}$ are given respectively by Eq. \eqref{dk} and \eqref{dkl}. 

Obviously, $\Delta_k$ is always positive for $E\leq 0$. However, it can be proved readily that $\Delta_{k,l}$ decreases monotonically by increasing $E$. Supposing the minimal solution to  $\Delta_{k,l}=0$ as $E_0 (<0)$, which can be obtained by setting $\omega_k=\omega_l=0$, one has 
\be
\Delta_{k,l}
\begin{cases}
  >0, & \mbox{if } E<E_0  \\
  \text{undetermined}, & \mbox{if }  E>E_0.
\end{cases}
\ee
Actually, $E_0$ is also a solution to Eq. \eqref{sbs3}, and thus depicts the single-excitation bound state energy. Directly, $\overline{c}_{1,k}$ becomes negative by Eq. \eqref{e33}  when $E< E_0$, as well as the integrand in Eq. \eqref{e3}.  Consequently, the solution can always be found for Eq. \eqref{e3} when $E<E_0$. 

When $E>E_0$, the positivity of $\Delta_{k,l}$ is undetermined because of the involvement of environment modes. Thus in this situation,  the calculation is strongly sensitive to the degree $n$ of orthogonal polynomials, and cannot attain stable results.

\section{Derivation of Eq. \eqref{e3}}\label{appendix:cklp}
To obtain Eq.\eqref{e3}, we first eliminate $c_{kkk}, c_{kkl}, c_{klp}$ in Eqs. \eqref{cklp},
\begin{widetext}
\be 
&\left(\Delta_{k, k}  + \frac{6 \left| g_k \right|^2}{E -3\omega_k}\right) c_{k, k}+ 2\sqrt{2} g_k c_{1,k} + 3\sqrt{2}g_k \sum_{i\left(\neq k \right)} \frac{g_l^*}{E- 2\omega_k -\omega_l} c_{k,l}=0, \nonumber \\
&\Delta_{k, l} c_{k, l}+ 2\left(g_k c_{1, l} + g_l c_{1, k} \right) + 3\sqrt{2}\left( \frac{g_k^* g_l c_{k, k}}{E- 2\omega_k -\omega_l} +  \frac{g_l^* g_k c_{l, l}}{E- 2\omega_l -\omega_k}\right)+ 3 g_l \sum_{p\left(\neq k \right)} \frac{g_p^* c_{k,p}}{E- \omega_k -\omega_l -\omega_p}+3 g_k \sum_{p\left(\neq l \right)} \frac{g_p^* c_{l,p}}{E- \omega_k -\omega_l -\omega_p} =0, \nonumber
\ee
\end{widetext}
in which
\be \label{dkl}
\Delta_{k, l}&=&\Delta + 2 \lambda + \omega_k +\omega_l+ 3 \sum_p \tfrac{\left| g_p \right|^2}{E - \omega_k -\omega_l -\omega_p} -E\nonumber \\
&=& \Delta + 2 \lambda + \omega_k +\omega_l+ 3 \int_{0}^{\infty}\text{d}\omega \tfrac{J(\omega)}{E- \omega_k -\omega_l-\omega} -E,\nonumber \\
\ee
For the second equality in Eq. \eqref{dkl},  Eq. \eqref{J} is applied.

Replacing $c_{k, k}$ by $\sqrt{2} c_{k, k}$, the two equations above can be rewritten as a single form,
\be \label{ckl}
&\Delta_{k, l} c_{k, l} + 2\left(g_k c_{1, l} + g_l c_{1, k} \right) \nonumber \\
& + 3 \sum_{p} \frac{g_p^*}{E- \omega_k -\omega_l -\omega_p}\left( g_l c_{k,p} +   g_k c_{l,p} \right)=0,
\ee
Eq. \eqref{ckl} combined with the first two equations in Eqs. \eqref{cklp} constitute a closed form for $c_0, c_{1,k}$ and $c_{k, l}$. 

To find correlation with $J(\omega)$,  we redefine $c_{1,k}, c_{k, l}$ as
\be
c_{1, k}&=& \sqrt{3}c_0 g_k \overline{c}_{1,k} , \\
c_{k, l}&=& \sqrt{3}c_0 g_k g_l \overline{c}_{k,l}. 
\ee 
Thus, one obtain respectively by Eq. \eqref{ckl} and the first two equations in Eqs. \eqref{cklp}, 
\be 
\begin{aligned}
&\left[ 3 \left(\Delta+ U \right) -E + 3 \sum_k \left|g_k \right|^2 \overline{c}_{1, k}\right] c_0=0,  \\
&\Delta_k \overline{c}_{1,k}  +1 + 2\sum_l  \left|g_l \right|^2 \overline{c}_{k,l}=0,   \\
&\Delta_{k,l} \overline{c}_{k,l} + \overline{c}_{1, k} + \overline{c}_{1, l}+ 3\sum_p \tfrac{\left|g_p \right|^2 }{E- \omega_k -\omega_l -\omega_p} \left( \overline{c}_{k, p} + \overline{c}_{l, p} \right)=0,   \\
\end{aligned}
\ee
in which
\be \label{dk}
\Delta_k= 2\left(\Delta+ \lambda \right) +U +\omega_k -E.
\ee
Finally, replacing $\sum_p  \left|g_p \right|^2$  by $\int_{0}^{\infty} \text{d}\omega_p$, one obtains Eq. \eqref{e3} and Eq. \ref{e33}

\end{document}